\documentclass[preprint,aps]{revtex4}

\newcommand{\cC}{\ensuremath{\mathcal{C}}}
\newcommand{\cP}{\ensuremath{\mathcal{P}}}
\newcommand{\cT}{\ensuremath{\mathcal{T}}}
\newcommand{\cQ}{\ensuremath{\mathcal{Q}}}

\begin{document}

\title{$\cal{P}\cal{T}$ symmetry in relativistic quantum mechanics}

\author{Carl~M.~Bender${}^1$ and Philip~D.~Mannheim${}^2$}

\affiliation{${}^1$Physics Department\\ Washington University\\ St.~Louis, MO
63130, USA\\ {\tt electronic address: cmb@wustl.edu}\\ \\
${}^2$Department of Physics\\ University of Connecticut\\ Storrs, CT 06269, USA
\\{\tt electronic address: philip.mannheim@uconn.edu}}

\date{2 July 2011}

\begin{abstract}
In nonrelativistic quantum mechanics and in relativistic quantum field theory,
time $t$ is a parameter and thus the time-reversal operator $\cT$ does not
actually reverse the sign of $t$. However, in relativistic quantum mechanics the
time coordinate $t$ and the space coordinates $\textbf{x}$ are treated on an
equal footing and all are operators. In this paper it is shown how to extend
$\cP\cT$ symmetry from nonrelativistic to relativistic quantum mechanics by
implementing time reversal as an operation that changes the sign of the time
coordinate operator $t$. Some illustrative relativistic quantum-mechanical
models are constructed whose associated Hamiltonians are non-Hermitian but $\cP
\cT$ symmetric, and it is shown that for each such Hamiltonian the energy
eigenvalues are all real.
\end{abstract}

\maketitle

\section{Introduction}
\label{s1}
In nonrelativistic quantum mechanics the position $\textbf{x}(t)$ is taken to be
an operator while the time $t$ is only a $c$-number parameter. To make quantum
mechanics relativistic one must treat time and space equivalently. There are
then two possibilities: one can either demote the spatial coordinates to
parameters or promote the time coordinate to an operator. The former
prescription is used in quantum field theory, where the field operators are
treated as functions of the spacetime parameters $\textbf{x}$ and $t$, but one
can also construct sensible quantum-mechanical theories via the latter approach
\cite{A1,A2}. In such theories a new parameter is needed to parameterize
evolution, and thus one introduces a fifth coordinate $\tau$ that is an
$SO(3,1)$ Lorentz scalar. In this five-dimensional formalism the space and
time coordinates $x^\mu(\tau)$ become operator functions of $\tau$ and one
obtains an $SO(3,1)$-invariant relativistic first-quantized generalization of
the nonrelativistic Heisenberg algebra $[x_j,p_k]=i\delta_{j,k}$:
\begin{equation}
[x^\mu(\tau),p^\nu(\tau)]=i\eta^{\mu\nu},\qquad
[x^\mu(\tau),p_\nu(\tau)]=i\delta^\mu_\nu,
\label{e1}
\end{equation}
where $\eta^{\mu\nu}$ is the $SO(3,1)$ Minkowski metric.

The dynamics in this formalism is $SO(3,1)$ invariant in the four operators
$x^\mu$, but is nonrelativistic in the fifth coordinate $\tau$ [because the
dynamics is not $SO(4,1)$ or $SO(3,2)$ invariant], and propagation is forward in
$\tau$. However, just as the nonrelativistic quantum-mechanical operator
$\textbf{x}(t)$
can propagate forward and backward with respect to its time parameter $t$, in
relativistic quantum mechanics all four components of $x^\mu(\tau)$ can
propagate forward and backward in $\tau$ \cite{A3}. The five-dimensional
formalism of \cite{A1,A2} readily incorporates forward and backward time
propagation, so one can introduce antiparticles with first quantization alone
without requiring the second-quantization techniques of quantum field theory.

When the five-dimensional Hamiltonian operator $\hat{H}$ is Hermitian and its
eigenfunctions have the separable form $\psi_n(x^\mu,\tau)=\phi_n(x^\mu) e^{-i
E_n\tau}$ and when its states obey the standard Dirac completeness relation
\begin{equation}
\sum|n\rangle\langle n|=I,
\label{e2}
\end{equation}
the five-space forward propagator takes the form
\begin{equation}
G_5(x^\mu_f,\tau;x^\mu_i,0)=-i\theta(\tau)\langle x_f^{\mu}|e^{-i\hat{H}\tau}
|x_i^\mu\rangle=-i\theta(\tau)\sum\phi_n(x_f^\mu)\phi_n^*(x_i^\mu)e^{-iE_n\tau}.
\label{e3}
\end{equation}
This propagator obeys a Schr\"odinger equation that is first order in $\tau$:
\begin{equation}
\left(i\partial_\tau+\hat{H}\right)G_5(x^\mu,\tau;0,0)=\delta(\tau)\delta^4(x^\mu).
\label{e4}
\end{equation}

The four-dimensional propagators in the five-dimensional formalism are
constructed by integrating out the fifth coordinate. Given (\ref{e3}), the
associated four-dimensional propagator is then defined as
\begin{equation}
G_4(x^\mu_f;x^\mu_i)=N\int_{-\infty}^{\infty}d\tau
G_5(x^\mu_f,\tau;x^\mu_i,0),
\label{e5}
\end{equation}
where $N$ is a normalization constant. Using the integral representation
\begin{equation}
\theta(\tau)=-\frac{1}{2\pi i}\int_{-\infty}^\infty d\nu\frac{e^{-i\nu\tau}}
{\nu+i\epsilon},
\label{e6}
\end{equation}
one then obtains
\begin{equation}
G_4(x^\mu_f;x^\mu_i)=N\sum\frac{\phi_n(x_f^\mu)\phi_n^*(x_i^\mu)}
{-E_n+i\epsilon}.
\label{e7}
\end{equation}
Finally, since the wave functions are eigenfunctions of $\hat{H}$,
$G_4(x^\mu_f;x^\mu_i)$ obeys
\begin{equation}
-\hat{H}G_4(x^\mu,0)=N\delta^4(x^\mu).
\label{e8a}
\end{equation}

The primary objective in using the approach of \cite{A1,A2} is to choose a
five-dimensional $\hat{H}$ so that $G_4(x^\mu,0)$ obeys a differential wave
equation of the form
\begin{equation}
D_4G_4(x^\mu,0)=\delta^4(x^\mu),
\label{e8}
\end{equation}
where $D_4$ is one of the familiar wave operators that appear in quantum field
theory (such wave operators are typically higher than first derivative in time).
[Normalizing $G_4(x^\mu,0)$ according to (\ref{e8}) would fix the constant $N$.]
Thus, in the simple case where the five-dimensional Hamiltonian has the form
$\hat{H}=\hat{\bar{p}}^2-(\hat{p}^0)^2+m^2$, (\ref{e7}) is the Fourier transform
of the standard four-dimensional scalar field Feynman propagator $D_4=
\partial_\mu\partial^\mu$. Because forward propagation in $\tau$ gives the
correct $i\epsilon$ prescription for the causal Feynman contour in four
dimensions, $D_4=\partial_\mu\partial^\mu$ is the usual four-dimensional
Klein-Gordon operator. Using the five-dimensional formalism, one can solve for a
one-body quantum-mechanical Schr\"odinger-type propagator in five space, and
from it one can construct a many-body quantum-field-theoretic propagator in four
space. The five-space formalism also permits one to choose five-space
Hamiltonians for which the resulting four-space propagator does not obey an
equation of the form (\ref{e8}) with a familiar $D_4$. In this paper we
construct some simple models that lead to a propagator equation with a familiar
$D_4$ and some that have a more general structure.

When Lorentz invariance was introduced in classical mechanics, it described the
invariance properties of the line element $ds^2=dt^2-d\textbf{x}^2$. In addition
to invariance under the continuous orthochronous Lorentz transformations, the
line element also possesses a set of discrete invariances, namely, space
reflection $\cP$: $\textbf{x}\to-\textbf{x}$, $t\to t$, time reversal $\cT$:
$\textbf{x}\to\textbf{x}$, $t\to-t$, and their product spacetime reflection $\cP
\cT$: $\textbf{x}\to-\textbf{x}$, $t\to-t$. However, when time reversal was
introduced into quantum mechanics by Wigner, the time reflection of the $i
\partial/\partial t$ operator was achieved not by replacing $t$ by $-t$ but
rather by taking $\cT$ to be an antiunitary operator that transforms $i$ into
$-i$ ($\cT: i\rightarrow-i$); time $t$ was treated as a $c$-number parameter
that is not affected by $\cT$. In relativistic quantum field theory, time
reversal is also not implemented by making the direct replacement $t\to-t$ even
though the line element $ds^2=dt^2-d\textbf{x}^2$ possesses this time-reversal
invariance. As noted above, in the five-dimensional relativistic
quantum-mechanical approach used here, we treat time as an operator and thus we
can implement a time-reversal operation that acts directly on the time. We can
also implement $\cP\cT$ transformations directly on the time operator.

The ability to implement a $\cP\cT$ transformation on the time operator is
appealing because of the implications of $\cP\cT$ invariance for Hamiltonians
that are not Hermitian. In the last few years it has been recognized
\cite{A4,A5,A6,A7} that a quantum-mechanical Hamiltonian that is not Hermitian
may still have an entirely real set of energy eigenvalues. In the cases that
were explicitly considered in \cite{A4,A5,A6,A7}, the reality of the eigenvalues
was traced to the existence of an underlying invariance of the Hamiltonian with
respect to a combined $\cP\cT$ reflection. Thus, while Dirac Hermiticity of the
Hamiltonian is sufficient for reality of eigenvalues, it is not necessary.  (Of
course, Hamiltonians with entirely real eigenvalues can be both $\cP\cT$
invariant and Dirac Hermitian.) However, recently it has been shown \cite{A8}
that a Hamiltonian that is not $\cP\cT$ invariant cannot have an entirely real
set of energy eigenvalues. This means that $\cP\cT$ invariance, in contrast to
Dirac Hermiticity, is necessary for the reality of energy eigenvalues \cite{A9}.
(If one knows only that a Hamiltonian is not Dirac Hermitian, one can say
nothing about the reality of the eigenvalues.) Thus, $\cP\cT$ invariance of a
Hamiltonian is a broader requirement than Dirac Hermiticity.

In the non-Hermitian $\cP\cT$-invariant context we apply the five-dimensional
formalism described above. To do this we recall \cite{A9} that when a
Hamiltonian is $\cP\cT$ invariant, its eigenvalues are either real or they come
in complex conjugate pairs. Consequently, both $\hat{H}$ and its Dirac-Hermitian
conjugate $\hat{H}^{\dagger}$ have the same eigenspectrum, and they are related
by some similarity transform $V$ \cite{A9a}
\begin{equation}
V\hat{H}V^{-1}=\hat{H}^{\dagger}.
\label{e9}
\end{equation}
In this case, if $|n\rangle$ is a right eigenvector $|R\rangle$ of $\hat{H}$,
then $\langle n|V$ rather then $\langle n|$ is a left eigenvector $\langle L|$
of $\hat{H}$. Consequently, the energy-eigenstate-completeness relation
(\ref{e2}) is replaced by
\begin{equation}
\sum|R\rangle\langle L|=\sum|n\rangle\langle n|V=I,
\label{e10}
\end{equation}
and (\ref{e3}) and (\ref{e7}) are replaced by
\begin{equation}
G_5(x^\mu_f,\tau;x^\mu_i,0)=-i\theta(\tau)\langle x_f^{\mu}|e^{-i\hat{H}\tau}
|x_i^{\mu}\rangle=-i\theta(\tau)\sum\langle x_f^{\mu}|n\rangle e^{-iE_n\tau}
\langle n|V|x_i^{\mu}\rangle,
\label{e11}
\end{equation}
\begin{equation}
G_4(x^\mu_f;x^\mu_i)=N\sum\frac{\langle x_f^{\mu}|n\rangle\langle
n|V|x_i^{\mu}\rangle}{-E_n+i\epsilon}.
\label{e12}
\end{equation}
The propagator (\ref{e12}) is the relevant one in the $\cP\cT$ case, and
with its $\cP\cT$-symmetric $\hat{H}$, it also obeys (\ref{e8a}).

Invariance under $\cP\cT$ reflection is a more physical requirement than
Hermiticity because the proper orthochronous Lorentz group has a complex $\cP
\cT$ extension. Until now, this aspect of the Lorentz group has not been
utilized because transformations that reverse the sign of the time have not
been considered. In the present paper we consider such transformations and
explicitly extend $\cP\cT$ symmetry to the relativistic quantum-mechanical
domain. In particular we study some simple non-Hermitian but $\cP\cT$-symmetric
$SO(3,1)$-invariant model Hamiltonians using the five-dimensional formalism and
for each Hamiltonian we show that all of the energy eigenvalues are real.

\section{A Simple Five-dimensional $\cP\cT$-symmetric Hamiltonian}
\label{s2}

The generic five-dimensional action has the form $I=\int_0^\tau d\tau'L(\tau')$,
where $\tau$ is the end point of integration. We begin with a simple example
that illustrates the five-dimensional formalism. Specifically, we take a
Lagrangian of the form
\begin{equation}
L=\frac{m}{2}\dot{x}_\mu\dot{x}^\mu-\frac{m\omega^2}{2}\left(x_\mu x^\mu
-2ia_\mu x^\mu-a_\mu a^\mu\right),
\label{e13}
\end{equation}
where $\mu=(0,1,2,3)$, the dot denotes differentiation with respect to $\tau$,
and $a^\mu$ is a real, external, $\tau$-independent four-vector operator
that commutes with $x^\mu$. As constructed, the action is a relativistic
$SO(3,1)$ scalar function of the four $x^\mu$ coordinates, but it is
nonrelativistic in the fifth coordinate $\tau$. We define a canonical momentum
\begin{equation}
p_{\mu}\equiv\frac{\delta I}{\delta\dot{x}^\mu}=m\dot{x}_\mu,
\label{e14}
\end{equation}
and then eliminate $\dot{x}^\mu$ to obtain a canonical Hamiltonian
\begin{eqnarray}
H&=&p_{\mu}\dot{x}^\mu-L\nonumber\\
&=&\frac{1}{2m}p_\mu p^\mu+\frac{m\omega^2}{2}\left(x_\mu x^\mu -2ia_\mu
x^\mu-a_\mu a^\mu\right).
\label{e15}
\end{eqnarray}
The Hamiltonian (\ref{e15}) is not Dirac Hermitian because of the $ia_\mu
x^\mu$ term.

Next, we assign $\cP$ and $\cT$ quantum numbers to the $x^\mu$ and $p^\mu$
operators, just as we do with the nonrelativistic $\textbf{x}$ and $\textbf{p}=d
\textbf{x}/dt$; to wit, we take the three spatial components $x^k$ to be $\cP$
odd [$\cP x^k(\tau)\cP^{-1}=-x^k(\tau)$] and $\cT$ even [$\cT x^k(\tau)\cT^{-1}=
x^k(-\tau)$], and take the three spatial components $p^k=dx^k/d\tau$ to be $\cP$
odd [$\cP p^k(\tau)\cP^{-1}=-p^k(\tau)$] and $\cT$ odd [$\cT p^k(\tau)\cT^{-1}=-
p^k(-\tau)$]. Similarly, we take the time component $x^0$ to be $\cP$ even [$\cP
x^0(\tau)\cP^{-1}=x^0(\tau)$] and $\cT$ odd [$\cT x^0(\tau)\cT^{-1}=-x^0(-
\tau)$] and take the time component $p^0=dx^0/d\tau$ to be $\cP$ even [$\cP p^0
(\tau)\cP^{-1}=p^0(\tau)$] and $\cT$ even [$\cT p^0(\tau)\cT^{-1}=p^0(-\tau)$].
With these assignments the four $x^\mu$ are $\cP\cT$ odd while the four $p^\mu$
are $\cP\cT$ even. Because $\cT$ also converts $i$ to $-i$, these assignments
are consistent with the commutation algebra in (\ref{e1}). We summarize these
assignments as follows:
\begin{equation}
\begin{array}{c|cccc} &\textbf{p}&p^0&\textbf{x}&x^0\\ \hline\cP&-&+&-&+\\
\cT&-&+&+&-\\ \cP\cT&+&+&-&-\\ \end{array}
\label{e16}
\end{equation}
Finally, we take the four-vector $a^\mu$ to be $\cP\cT$ even. For our purposes
we will need $\textbf{a}$ to be $\cP$ even and thus $\cT$ even, and $a^0$ to be
$\cP$ odd and thus $\cT$ odd. In the five-space the Hamiltonian (\ref{e15}) is
conjugate to $\tau$ and not to $x^0$. Neither $\cP$ nor $\cT$ affect $\tau$
because $\tau$ is only a parameter, so with $p_\mu\dot{x}^\mu =p_\mu
p^\mu/m$ being $\cP\cT$ even, the Hamiltonian is $\cP\cT$ symmetric.

To determine the energy eigenvalues we take the spacetime metric to be
$\textrm{diag}(\eta_{\mu\nu})=(-1,1,1,1)$. Writing $x^\mu=(t,x,y,z)$, we obtain
a wave-mechanics representation of the algebra (\ref{e1}) when $p_\mu=-i\partial
/\partial x^\mu$; that is,
\begin{equation}
p_0=-i\frac{\partial}{\partial t},\qquad p_k=-i\frac{\partial}{\partial x^k}.
\label{e17}
\end{equation}
Consequently, in five-space the Schr\"odinger equation takes the form
\begin{equation}
i\frac{\partial\psi(\tau,x^\mu)}{\partial\tau}=\left[-\frac{1}{2m}\eta^{\mu\nu}
\frac{\partial}{\partial x^\mu}\frac{\partial}{\partial x^\nu}+\frac{m\omega^2}
{2}(x_\mu-ia_\mu)(x^\mu-ia^\mu)\right]\psi(\tau,x^\mu).
\label{e18}
\end{equation}
The substitution $y^\mu=x^\mu-ia^\mu$ brings (\ref{e18}) to the form
\begin{equation}
i\frac{\partial\psi(\tau,y^\mu)}{\partial\tau}=\left[\frac{1}{2m}\left(\frac{
\partial^2}{\partial t^2}-\frac{\partial^2}{\partial\textbf{y}^2}\right)+\frac
{m\omega^2}{2}(\textbf{y}^2-t^2)\right]\psi(\tau,y^\mu),
\label{e19}
\end{equation}
and reduces the Schr\"odinger equation to a four-dimensional harmonic
oscillator with Minkowski signature. Noting that
\begin{equation}
\left[\frac{1}{2m}\left(\frac{\partial^2}{\partial t^2}-\frac{\partial^2}{
\partial\textbf{y}^2}\right)+\frac{m\omega^2}{2}(\textbf{y}^2-t^2)\right]
e^{-m\omega(\textbf{y}^2-t^2)/2}=2\omega e^{-m\omega(\textbf{y}^2-t^2)/2},
\label{e20}
\end{equation}
we see that the $t$-dependent sector contributes a positive zero-point energy
equal to $\omega/2$ just as the $\textbf{y}$-dependent sector does. Because all
the eigenvalues of a harmonic oscillator are real, the five-space energy
eigenvalues of (\ref{e18}) are given by
\begin{equation}
E_5=(n_x+n_y+n_z+n_t+2)\omega,
\label{e21}
\end{equation}
where each of $n_x$, $n_y$, $n_z$ and $n_t$ ranges over the positive integers.
Thus, while the Hamiltonian (\ref{e15}) is not Hermitian, all of its energy
eigenvalues are real.

For this model the five-space propagator obeys
\begin{equation}
\left[i\frac{\partial}{\partial\tau}+\frac{1}{2m}\frac{\partial}{\partial
x_\mu}\frac{\partial}{\partial x^\mu}-\frac{m\omega^2}{2}(x_\mu-ia_\mu)(x^\mu
-ia^\mu)\right]G_5(x^\mu,\tau;0,0)=\delta(\tau)\delta^4(x^\mu)
\label{e22}
\end{equation}
and we show in Appendix A that
\begin{equation}
G_5(x^\mu,\tau;0,0)=\theta(\tau)\frac{1}{(\sin\omega\tau)^2}\exp\left[\frac{im
\omega\cos\omega\tau(x_\mu-ia_\mu)(x^\mu-ia^\mu)}{2\sin\omega\tau}\right].
\label{e23}
\end{equation}
The propagator of the associated four-dimensional theory is then obtained via
(\ref{e5}), and it obeys (\ref{e8a}) with the $\cP\cT$-symmetric
$\hat{H}=-\partial_\mu\partial^\mu/2m+m\omega^2(x_\mu-ia_\mu)(x^\mu-ia^\mu)/2$.

Using the $\cP\cT$-theory techniques described in \cite{A6}, one can
demonstrate
the reality of the eigenvalues algebraically without actually solving the
Schr\"odinger equation. To do so, one must construct an operator $e^{\cQ}$ that
possesses four key properties: (i) a similarity transformation using $e^\cQ$
preserves the commutation relations; (ii) $\cQ$ is a Hermitian operator (so that
$e^{\cQ}$ is not unitary); (iii) like $V$ in (\ref{e9}), $e^{\cQ}$ effects the
transformation
\begin{equation}
e^{-\cQ}He^\cQ =H^\dagger;
\label{e24}
\end{equation}
(iv) the operator
\begin{equation}
\tilde H=e^{-\cQ/2}He^{\cQ/2}
\label{e25}
\end{equation}
obeys $\tilde{H}^\dag=\tilde{H}$. The existence of such a $\cQ$ operator
implies that the energy eigenvalues of $H$ are all real.

We now construct the $\cQ$ operator for our simple five-dimensional model.  Note
that the momentum operator will effect the transformation
\begin{equation}
e^{-b^\nu p_\nu}x^\mu e^{b^\rho p_\rho}=x^\mu+ib^\mu,
\label{e26}
\end{equation}
and leave the commutation relations (\ref{e1}) untouched for any four-vector
$b^\mu$ that commutes with both $x^\mu$ and $p^\mu$. Given (\ref{e26}), we
identify $\cQ$ as the Hermitian operator $2a^\nu p_\nu$ because
\begin{equation}
e^{-2a^\nu p_\nu}H e^{2a^\rho p_\rho}=\frac{1}{2m}p_\mu p^\mu+\frac{m\omega^2}
{2}\left(x_\mu x^\mu +2ia_\mu x^\mu-a_\mu a^\mu\right)=H^{\dagger}.
\label{e27}
\end{equation}
Similarly, the transformation
\begin{equation}
e^{-a^\nu p_\nu}H e^{a^\rho p_\rho}=\frac{1}{2m}p_\mu p^\mu+\frac{m\omega^2}{2}
x_\mu x^\mu=\tilde{H}
\label{e28}
\end{equation}
generates an equivalent Hamiltonian $\tilde{H}$ that is manifestly Hermitian.

In $\cP\cT$ quantum mechanics one introduces an operator $\cC$ that is required
to obey
\begin{equation}
[\cC,H]=0,\qquad\cC^2=I.
\label{e29}
\end{equation}
One constructs this operator by making the {\it ansatz} $\cC=e^\cQ\cP$, where
the operator $\cP$ obeys $\cP^2=I$. In this form, the operator $\cC$ fulfills
the condition $\cC^2=I$ provided that $\cQ$ satisfies $\cP\cQ\cP=-\cQ$. With
$e^{-\cQ}$ generating $e^{-\cQ}He^\cQ=H^\dagger$, the operator $\cC$ obeys $\cC^
{-1}H\cC=H$ if $\cP$ generates $\cP H\cP=H^\dagger$. For the $\cQ$ and $H$ of
interest here, both $\cP\cQ\cP=-\cQ$ and $\cP H\cP=H^\dagger$ hold provided that
$a^0$ is $\cP$ odd and $\textbf{a}$ is $\cP$ even. With this choice for the
parity of $a^\mu$, we then identify $\cC=e^\cQ\cP$. (Previously, we had required
that $a^\mu$ be $\cP\cT$ even.) Then, if both $a^\mu$ and $p^\mu$ are $\cP\cT$
even, the operator $\cQ$ is $\cP\cT$ even. As constructed, $\cC$ thus obeys
$[\cC,\cP\cT]=0$, as expected \cite{A8,A9} when all energy eigenvalues are real
\cite{A10a}.

\section{Five-dimensional Pais-Uhlenbeck Oscillator}
\label{s3}

In 1950 Pais and Uhlenbeck \cite{A11} explored the question of whether the
Pauli-Villars regulator associated with the fourth-order equation of motion
\begin{equation}
(\partial_t^2-\nabla^2+M_1^2)(\partial_t^2-\nabla^2+M_2^2)\phi(\textbf{x},t)=0
\label{e30}
\end{equation}
and propagator
\begin{equation}
D(k^2)=\frac{1}{(k^2+M_1^2)(k^2+M_2^2)}=\frac{1}{M_2^2-M_1^2}\left(\frac{1}{k^2+
M_1^2}-\frac{1}{k^2+M_2^2}\right),
\label{e31}
\end{equation}
where $k^2=-(k^0)^2+\textbf{k}^2$, could be physically viable, or whether it was
merely a mathematical technique to regulate Feynman integrals. To this end
they replaced the scalar field $\phi(\textbf{x},t)$ by a single coordinate $z
(t)$ and examined single momentum modes $\omega_1^2=\textbf{k}^2+M_1^2$ and
$\omega_2^2=\textbf{k}^2+M_2^2$. The resulting nonrelativistic
quantum-mechanical limit of the equation of motion (\ref{e30}) and the
propagator (\ref{e31}),
\begin{equation}
(\partial_t^2+\omega_1^2)(\partial_t^2+\omega_2^2)z(t)=0,\qquad G(E)=\frac{1}
{\omega_1^2-\omega_2^2}\left(\frac{1}{E^2-\omega_1^2}-\frac{1}{E^2-\omega_2^2}
\right),
\label{e32}
\end{equation}
is known as the PU oscillator.

Pais and Uhlenbeck found that if the theory were quantized with a standard
positive-metric Hilbert space, the energy spectrum would not be bounded below.
One can evade this negative-energy problem by quantizing the theory in a
negative-metric Hilbert space, but as the relative minus sign in (\ref{e32})
indicates, the disadvantage of doing so is that one obtains states of negative
Dirac norm and evidently loses unitarity.

The PU oscillator was revisited in 2008 \cite{A12,A13} and a new realization of
the theory was found in which the Hilbert space has neither negative-energy nor
negative-norm states. In this realization the Hamiltonian is not Dirac-Hermitian
but is instead $\cP\cT$ invariant. The norm is given by $\langle L|R\rangle=
\langle n|V|n\rangle$, rather than by the Dirac norm $\langle n|n\rangle$, and
the completeness relation is given by (\ref{e10}) rather than by (\ref{e2}). In
analogy with (\ref{e12}), the relative minus signs in (\ref{e31}) and
(\ref{e32}) are generated by the presence of the $V$ operator in the propagator
and not by quantizing with an indefinite metric. This realization took a long
time (more than half a century) to discover because the Hamiltonian of the
theory appeared to be Dirac Hermitian even though it is not. (In
Refs.~\cite{A12,A13} the nonrelativistic $\cP\cT$ realization of the PU
oscillator is studied, and in Ref.~\cite{A13} the relativistic scalar field
theory is examined.)

For the case of the nonrelativistic PU oscillator, the equation of motion
(\ref{e32}) for the coordinate $z(t)$ can be derived by a stationary variation
of the PU oscillator action
\begin{equation}
I_{\rm PU}=\frac{\gamma}{2}\int dt\left[{\ddot z}^2-\left(\omega_1^2+\omega_2^2
\right){\dot z}^2+\omega_1^2\omega_2^2z^2\right],
\label{e33}
\end{equation}
where $\gamma$, $\omega_1$ and $\omega_2$ are positive constants. Since
$\dot{z}$ serves as the conjugate of both $z$ and $\ddot{z}$, the action is
constrained. One thus replaces $\dot{z}$ by a new variable $x$, and using the
method of Dirac constraints, one obtains \cite{A14,A15} the Hamiltonian
\begin{equation}
H_{\rm PU}=\frac{p_x^2}{2\gamma}+p_zx+\frac{\gamma}{2}\left(\omega_1^2+
\omega_2^2\right)x^2-\frac{\gamma}{2}\omega_1^2\omega_2^2z^2
\label{e34}
\end{equation}
with two canonical pairs that obey $[x,p_x]=i$ and $[z,p_z]=i$.

In the realization of the theory for which the energy eigenvalues are bounded
below, $H_{\rm PU}$ appears to be Hermitian but it is not. Specifically, one
solves the Schr\"odinger equation for the ground state of the system with energy
$E_0=(\omega_1+\omega_2)/2$. The eigenfunction is
\begin{equation}
\psi_0(z,x)={\rm
exp}\left[\frac{\gamma}{2}(\omega_1+\omega_2)\omega_1\omega_2
z^2+i\gamma\omega_1\omega_2zx-\frac{\gamma}{2}(\omega_1+\omega_2)x^2\right].
\label{e35}
\end{equation}
This eigenfunction diverges exponentially for large $z$, so integration by parts
generates surface terms that cannot be discarded. Thus, one cannot represent the
operator $p_z$ by $-i\partial_z$. However, one can replace $z$ by $iz$ (this
is equivalent to working in a Stokes wedge in the complex-$z$ plane that
includes the imaginary $z$ axis but not the real one \cite{A12}), and represent
$p_z$ by $-i\partial_{iz}=-\partial_z$. The eigenfunction then vanishes
exponentially as $z$ becomes large. The highly unusual implication of the
structure of (\ref{e35}) (and the reason it took so long to find) is that while
both conjugate pairs of coordinates are obtained from the same Lagrangian, the
commutator $[x,p_x]=i$ is realized by Hermitian operators, while the commutator
$[z,p_z]=i$ is realized by anti-Hermitian operators. As a result, the $p_zx$
cross-term in (\ref{e34}) is not Hermitian, and the Hamiltonian $H_{\rm PU}$ is
also not Hermitian.

Rather than using non-Hermitian operators, we make the similarity transformation
\begin{equation}
y=e^{\pi p_zz/2}ze^{-\pi p_zz/2}=-iz,\qquad q=e^{\pi p_zz/2}p_ze^{-\pi p_zz/2}=
ip_z,
\label{e36}
\end{equation}
to construct Hermitian operators $y$ and $q$ that obey $[y,q]=i$. In terms of
$y$ and $q$ the Hamiltonian now takes the form
\begin{equation}
H_{\rm PU}=\frac{p^2}{2\gamma}-iqx+\frac{\gamma}{2}\left(\omega_1^2+\omega_2^2
\right)x^2+\frac{\gamma}{2}\omega_1^2\omega_2^2y^2,
\label{e37}
\end{equation}
where for notational simplicity we have replaced $p_x$ by $p$. The Hamiltonian
$H_{\rm PU}$ is now manifestly non-Hermitian.

While $H_{\rm PU}$ is not Hermitian, the $\cP$ and $\cT$ quantum-number
assignments
\begin{equation}
\begin{array}{c|cccc} &p&x&q&y\\ \hline\cP&-&-&+&+\\ \cT&-&+&+&-\\
\cP\cT&+&-&+&-\\ \end{array}
\label{e38}
\end{equation}
make $H_{\rm PU}$ symmetric under $\cP\cT$ reflection. Introducing the operator
\begin{equation}
\cQ=\alpha pq+\beta xy,\qquad \alpha=\frac{1}{\gamma\omega_1\omega_2}\log\left(
\frac{\omega_1+\omega_2}{\omega_1-\omega_2}\right),\qquad\beta=\alpha\gamma^2
\omega_1^2\omega_2^2,
\label{e39}
\end{equation}
we then find that \cite{A12,A13} the similarity-transformed PU Hamiltonian
\begin{equation}
\tilde{H}_{\rm PU}=e^{-\cQ/2}H_{\rm PU}e^{\cQ/2}=\frac{{p}^2}{2\gamma}+\frac{
q^2}{2\gamma\omega_1^2}+\frac{\gamma}{2}\omega_1^2 x^2+\frac{\gamma}{2}
\omega_1^2\omega_2^2{y}^2
\label{e40}
\end{equation}
represents two uncoupled harmonic oscillators. The transformed Hamiltonian
$\tilde{H}_{\rm PU}$ in (\ref{e40}) is both Hermitian and manifestly positive
definite. This realization of the quantum theory, which is associated with the
non-Hermitian $H_{\rm PU}$, has no negative-norm or negative-energy eigenstates
\cite{A16}.

Because the transformation with $e^{\cQ/2}$ is not unitary, the propagator
\begin{equation}
D(H_{\rm PU})=\langle x^{\prime},y^{\prime}|e^{-iH_{\rm PU}t}|x,y\rangle=\langle
x^{\prime},y^{\prime}|e^{\cQ/2}e^{-i\tilde{H}_{\rm PU}t}e^{-\cQ/2}|x,y\rangle
\label{e41}
\end{equation}
associated with $H_{\rm PU}$ does not transform into the propagator
\begin{equation}
D(\tilde{H}_{\rm PU})=\langle x^\prime,y^\prime|e^{-i\tilde{H}_{\rm PU}t}|x,y
\rangle
\label{e42}
\end{equation}
that one would ordinarily associate with a two-uncoupled-oscillator system. The
state $\langle x,y|e^{\cQ/2}$ is not the conjugate of $e^{-\cQ/2}|x,y\rangle$,
and the propagators in (\ref{e41}) and (\ref{e42}) are not equivalent; for this
realization of the PU Hamiltonian we must use (\ref{e41}) and not (\ref{e42}).
The dependence on the operator $V=e^{-\cQ}$ is crucial because it
generates the relative minus sign in (\ref{e32}).

We now illustrate $\cP\cT$ invariance in relativistic quantum mechanics by
applying the five-dimensional formalism to the PU oscillator. We will see that a
straightforward covariant generalization of the PU oscillator does not
lead back to (\ref{e31}). Consequently, in the next section we provide an
alternate five-dimensional formalism that does.

To generalize the PU oscillator to relativistic quantum mechanics we replace
(\ref{e33}) by
\begin{equation}
I=\frac{\gamma}{2}\int_0^{\tau}d\tau\left[\ddot{z}_{\mu}\ddot{z}^{\mu}-\left(
M_1^2+M_2^2\right)\dot{z}_{\mu}\dot{z}^{\mu}+M_1^2M_2^2z_{\mu}z^{\mu}\right],
\label{e43}
\end{equation}
where the dot denotes differentiation with respect to $\tau$. Because of
constraints associated with this action, the Hamiltonian has the form
\begin{equation}
H=\frac{(p_x)_\mu(p_x)^\mu}{2\gamma}+(p_z)_\mu x^\mu+\frac{\gamma}{2}\left(M_1^2
+M_2^2\right)x_\mu x^\mu-\frac{\gamma}{2}M_1^2M_2^2z_\mu z^\mu.
\label{e44}
\end{equation}

Recalling the transformation in (\ref{e36}), we let $y^\mu=-iz^\mu$ and $q^\mu=
i(p_z)^\mu$. On setting $(p_x)^\mu=p^\mu$ we obtain two canonical pairs of
operators that obey
\begin{equation}
[x^\mu(\tau),p^\nu(\tau)]=i\eta^{\mu\nu},\qquad
[q^\mu(\tau),y^\nu(\tau)]=i\eta^{\mu\nu},
\label{e45}
\end{equation}
and a Hamiltonian of the form
\begin{equation}
H=\frac{p_\mu p^\mu}{2\gamma}-iq_\mu x^\mu+\frac{\gamma}{2}\left(M_1^2
+M_2^2 \right)x_\mu x^\mu+\frac{\gamma}{2}M_1^2M_2^2y_\mu y^\mu.
\label{e46}
\end{equation}
The assignments
\begin{equation}
\begin{array}{c|cccccccc}
&\textbf{p}&p^0&\textbf{x}&x^0&\textbf{q}&q^0&\textbf{y}&y^0\\ \hline \cP&-&+&-&
+&+&-&+&-\\ \cT&-&+&+&-&+&-&-&+\\ \cP\cT&+&+&-&-&+&+&-&-\\ \end{array}
\label{e47}
\end{equation}
in which $x^0$ changes sign under $\cT$, then establish that $H_{\rm PU}$ is
$\cP\cT$ symmetric.

Next, we introduce the operator
\begin{equation}
\cQ=\alpha p_\mu q^\mu+\beta x_\mu y^\mu,\qquad \alpha=\frac{1}{\gamma M_1M_2}
\log\left(\frac{M_1+M_2}{M_1-M_2}\right),\qquad\beta=\alpha\gamma^2M_1^2M_2^2,
\label{e48}
\end{equation}
and find that
\begin{equation}
\tilde{H}=e^{-\cQ/2}He^{\cQ/2}=\frac{p_\mu p^\mu}{2\gamma}+\frac{q_\mu q^\mu}{2
\gamma M_1^2}+\frac{\gamma}{2}M_1^2 x_\mu x^\mu+\frac{\gamma}{2}M_1^2M_2^2
y_\mu y^\mu.
\label{e49}
\end{equation}
Thus, the energy eigenvalues of the $\cP\cT$-symmetric Hamiltonian $H$ are all
real.

We show in Appendix A that if we set $M_1=M$ and $M_2=0$, the five-space
propagator is
\begin{equation}
G_5(x^\mu,y^\mu,\tau;0,0,0)=\theta(\tau)\frac{e^{iB/A}}{A^2},
\label{e50}
\end{equation}
where
\begin{eqnarray}
2B/\gamma&=&Mx_\mu x^\mu(\sin M\tau-M\tau\cos M\tau)-M^3y_\mu y^\mu \sin M\tau
+2iM^2x_\mu y^\mu (1-\cos M \tau),\nonumber\\ A&=&2-2\cos M\tau-M\tau\sin M\tau.
\label{e51}
\end{eqnarray}

The propagator of the associated four-dimensional theory may now be obtained by
performing the integral in (\ref{e5}), and the resulting propagator will obey
(\ref{e8a}) with $\hat{H}=-(1/2\gamma)\partial/\partial x_\mu\partial/\partial
x^\mu-x^\mu\partial/\partial y^\mu+\gamma M^2x_\mu x^\mu/2$. While of
interest in itself, this propagator is not  of the generic Pauli-Villars
form given in (\ref{e31}). Thus, in Sec.~\ref{s4} we provide an alternate
choice for the five-dimensional Hamiltonian that will lead to (\ref{e31}).

\section{Alternate Formulation of the Five-space PU Oscillator}
\label{s4}

Given the structure of (\ref{e30}) we take the five-space $\hat{H}$ to have the
operator form
\begin{equation}
\hat{H}=-[-(\hat{p}^0)^2+\hat{\bar{p}}^2+M_1^2][-(\hat{p}^0)^2+\hat{\bar{p}}^2+
M_2^2].
\label{e52}
\end{equation}
For this Hamiltonian the five-dimensional energies are given by
\begin{equation}
E_5=-[-(p^0)^2+\bar{p}^2+M_1^2][-(p^0)^2+\bar{p}^2+M_2^2],
\label{e53}
\end{equation}
where the momenta in (\ref{e53}) are the eigenvalues of the operators in
(\ref{e52}). Inserting these energies into (\ref{e7}), we obtain the
Pauli-Villars propagator in (\ref{e31}), with (\ref{e8}) being satisfied.

Equation (\ref{e52}) leads directly to (\ref{e31}), but its use here is
nonstandard because it does not have a simple Lagrangian counterpart. In the
previous examples and in the derivation of the Klein-Gordon propagator, one can
start with a five-dimensional action (of the form $\int_0^{\tau}d\tau\,
\dot{x}_\mu\dot{x}^\mu$ for the specific Klein-Gordon case) and by a canonical
procedure derive a Hamiltonian from it. The Lagrangians in these examples are
quadratic functions of the coordinates, so the procedure is straightforward and
yields Hamiltonians that are also quadratic. However, the Hamiltonian
(\ref{e52}) is not quadratic; it is quartic  because the wave operator in
(\ref{e30}) is a fourth-order derivative operator \cite{A16a}. Since the
Lagrangian is given by $L(\dot{x}^\mu)=p_\mu \dot{x}^\mu-H(p_\mu p^{\mu})$ and
since $p_\mu=\partial L/\partial \dot{x}^\mu$, one can in principle construct
$L(\dot{x}^\mu)$ if one knows $H(p_\mu p^{\mu})$. Doing so for (\ref{e52}) is
difficult, so we start directly with $H(p_\mu p^{\mu})$. Once we have $H(p_\mu
p^\mu)$, we can then use the representation in (\ref{e7}) without needing to
know the structure of the Lagrangian.

We can recover the four-dimensional Pauli-Villars propagator, but at first it
appears that the Hamiltonian in (\ref{e52}) is Hermitian. Moreover, in the
second-order Klein-Gordon case with $\hat{H}=-(p^0)^2+\bar{p}^2+M_1^2$ and
real $E_5$ the Hamiltonian is Hermitian. However, in the fourth-order case, we
note that $(p^0)^2$ is given as
\begin{equation}
(p^0)^2=\frac{1}{2}\left(E_1^2+E_2^2\pm[(E_1^2-E_2^2)^2-4E_5]^{1/2}\right),
\label{e54}
\end{equation}
where $E_i^2=\bar{p}^2+M_i^2$. Thus, now there can be real values of $E_5$ for
which $(p^0)^2$ is complex and for which the operator $(\hat{p}^0)^2$, and thus
$\hat{H}$, is not Hermitian. (Note that with $E_5$ being real, the Hamiltonian
must be $\cP\cT$ invariant.)

In addition, we note that for general $M_1$ and $M_2$, if we take $E_5$ to be
zero, the eigenfunctions associated with the operator $\hat{H}$ in (\ref{e52})
will have the form $\psi_1=e^{-iE_1 t+i\bar{p}\cdot \bar{x}}$ and $\psi_2=e^{-i
E_2t+i\bar{p}\cdot\bar{x}}$. However, if we then set $M_1=0$ and $M_2=0$, there
will be eigenfunctions of the form $\psi_a=e^{-ipt+i\bar{p}\cdot\bar{x}}$ and
$\psi_b=n_\mu x^\mu e^{-ipt+i\bar{p}\cdot\bar{x}}$, where $n^\mu$ is the unit
timelike vector $n^\mu=(1,0,0,0)$. Of the two $\psi_a$ and $\psi_b$
eigenfunctions, only $\psi_a$ is stationary; $\psi_b$ grows linearly in the time
coordinate, which indicates that the Hamiltonian has Jordan-block form and that
it has an incomplete set of eigenvectors. Consequently, the Hamiltonian $\hat{H
}$ in (\ref{e52}) cannot be diagonalized and is not Hermitian. Since $\hat{H
}$ is not Hermitian when $E_5=M_1=M_2=0$, it must also not be Hermitian for a
range of values of these parameters. In Ref.~\cite{A13} it was found that in the
equal-frequency limit $\omega_1=\omega_2$ of the PU oscillator, the Hamiltonian
in (\ref{e37}) is also nondiagonalizable and non-Hermitian.

The solutions to (\ref{e54}) thus break up into two sectors. In one sector
the Hamiltonian is Hermitian and the energy eigenvalues are unbounded below
($4E_5<(M_1^2-M_2^2)^2$). In the other sector the Hamiltonian is not Hermitian
and the energies are bounded below ($4E_5>(M_1^2-M_2^2)^2$), just as in the case
of the nonrelativistic PU oscillator. If $E_5$ is real, the Hamiltonian is $\cP
\cT$ invariant in both cases. In the sector where $\hat{H}$ is Hermitian the
four-space propagator is given by (\ref{e7}). In the non-Hermitian sector the
four-space propagator is given by (\ref{e12}) and as before, the $V$ operator
then generates the relative minus sign in the Pauli-Villars propagator
\cite{A17}. Our five-space treatment of the Pauli-Villars propagator based on
(\ref{e52}) recovers the key features of the analyses of Refs.~\cite{A12,A13}.
We see that one can extend $\cP\cT$ symmetry to the five-dimensional formalism,
and while we have not directly studied the time-reversal and $\cP\cT$ properties
of the time operator in the Pauli-Villars case, those properties follow directly
from the commutation relations (\ref{e1}) depending on how they are explicitly
specified for $(\hat{p})^0$.

\section{Summary}

Using a number of elementary models, we have shown in this paper that the
standard techniques of $\cP\cT$ quantum mechanics extend and apply to
relativistic quantum mechanics, where the time-reversal operator $\cT$ reverses
the sign of the time operator $x^0$. We conclude that relativistic $\cP
\cT$-symmetric quantum mechanics is physically viable.

The work of CMB is supported by a grant from the U.S.~Department of Energy.

\appendix
\setcounter{equation}{0}
\def\theequation{A\arabic{equation}}

\section{Construction of Five-space Propagators}

To construct propagators that obey the five-dimensional equation $\left(i
\partial_\tau+\hat{H}\right)G_5(x^\mu,\tau;0,0)=\delta(\tau)\delta^4(x^\mu)$,
we first
recall how a propagator is constructed when the eigenmodes of $\hat{H}$ are
plane waves. For the nonrelativistic quantum-mechanical free particle in one
space dimension there is a plane wave basis and the propagator is given by
\begin{equation}
G_1(x,t;0,0)=-\frac{i\theta(t)}{2\pi}\int dp\,e^{-ipx-ip^2t/m}.
\label{F1}
\end{equation}
When $i\partial_t$ acts on $-i\theta(t)$, we generate the $\delta(t)\delta(x)$
term, while if we omit the $\theta(t)$ function, the rest of the propagator
obeys
\begin{equation}
\left[i\frac{\partial}{\partial t}+\frac{1}{2m}\frac{\partial^2}{\partial
x^2}\right]R_1(x,t;0,0)=0,
\label{F2}
\end{equation}
where $G_1(x,t;0,0)=\theta(t)R_1(x,t;0,0)$. The Fourier transform in
(\ref{F1})
can be performed analytically and yields
\begin{equation}
G_1(x,t;0,0)=\theta(t)\left(\frac{m}{2\pi i t}\right)^{1/2}e^{imx^2/2t}.
\label{F3}
\end{equation}
The term $I_{\rm STAT}=mx^2/2t$ in the exponent is the value of the classical
action $I=(m/2)\int_0^t dt\,\dot{x}^2$ for the stationary path $\ddot{x}=0$
between the end points $(x=0,t=0)$ and $(x,t)$.

If one were to calculate this propagator as a path integral $\int [dx]e^{iI}$
over a complete basis of paths between the end points, one would obtain the same
$e^{iI_{\rm STAT}}$ phase, but one would not know the multiplicative pre-factor.
This pre-factor is determined by requiring that the propagator obey (\ref{F2}).
(If one does not have a plane-wave basis, one can evaluate the propagator via a
path integral and then use the Schr\"odinger equation to determine the
pre-factor.)

For the one-dimensional harmonic oscillator (where the basis is not plane
waves), the path integral again has the form $e^{iI_{\rm STAT}}$, where $I_{\rm
STAT}$ is the value of $I=(m/2)\int_0^{T} dt[\dot{x}^2-\omega^2x^2]$ as
evaluated in the stationary path $\ddot{x}+\omega^2x=0$ between the end points
$(x=0,t=0)$ and
$(x=x_f,t=T)$. Noting that $\dot{x}^2-\omega^2x^2=d(x\dot{x})/dt-x\ddot{x}-
\omega^2x^2$, we obtain $I_{\rm STAT}=mx_f\dot{x}_f/2$. The solution to the
equation of motion is $x(t)=x_f\sin\omega t/\sin\omega T$, $\dot{x}(t)=\omega
x_f\cos\omega t/\sin\omega T$, so we obtain $I_{\rm STAT}=m\omega x_f^2\cos
\omega T/2\sin\omega T$. With this form for $I_{\rm STAT}$, the pre-factor
evaluates to $(\sin\omega T)^{-1/2}$ and the propagator is
\begin{equation}
G_1(x,T;0,0)=\theta(T)\left(\frac{1}{\sin \omega T}\right)^{1/2}\exp\left(
\frac{im\omega x^2\cos \omega T}{2\sin\omega T}\right).
\label{F4}
\end{equation}
The propagator (\ref{e23}) is the shifted covariant generalization of this
result.

The propagator associated with the PU oscillator action given in (\ref{e33}) has
already been reported in the literature \cite{A19}, and because the action is
quadratic, the $\int d[z]$ path integral between end points with fixed $z$ and
$\dot{z}$ has the form $\exp(iI_{\rm STAT})$ with the appropriate $I_{\rm
STAT}$. Here, we present a simplified version of the propagator in which we set
$\omega_1=\omega$, $\omega_2=0$. In this case the classical action reduces to
\begin{equation}
I_{\rm PU}=\frac{\gamma}{2}\int dt\left(\ddot{z}^2-\omega^2{\dot z}^2\right),
\label{F5}
\end{equation}
and the stationary classical equation of motion is given by
\begin{equation}
\partial_t^2(\ddot{z}+\omega^2z)=0.
\label{F6}
\end{equation}

Noting that
\begin{equation}
\partial_t\left(\dot{z}\ddot{z}-z\partial^3_tz-\omega^2z\dot{z}\right)
=\ddot{z}^2-\omega^2\dot{z}^2-z\partial_t^2(\ddot{z}+\omega^2z),
\label{F7}
\end{equation}
on evaluating $I_{\rm STAT}$ between $z=0$, $\dot{z}=0$ at $t=0$, and
$z(T)$, $\dot{z}(T)$ at $t=T$, we obtain
\begin{equation}
I_{\rm
STAT}=(\gamma/2)\left(\dot{z}(T)\ddot{z}(T)-z(T)\partial^3_tz(T)-\omega^2
z(T)\dot{z}(T)\right).
\label{F8}
\end{equation}
Hence, introducing
\begin{eqnarray}
\omega\alpha A(T)&=&\dot{z}(T)(\omega T-\sin\omega T)-\omega z(T)(1-\cos\omega
T), \nonumber\\
\beta A(T)&=&\dot{z}(T)(1-\cos\omega T)-z(T)\omega \sin \omega T, \nonumber\\
A(T)&=&2-2\cos \omega T-\omega T\sin \omega T,
\label{F9}
\end{eqnarray}
we find that the solution to (\ref{F6}) that satisfies the boundary conditions
takes the form
\begin{eqnarray}
z(t)&=&-\alpha(1-\cos \omega t )-(\beta/\omega)\sin\omega t+\beta t,\nonumber\\
\dot{z}(t)&=&-\alpha\omega\sin\omega t-\beta\cos \omega t +\beta,\nonumber\\
\ddot{z}(t)&=&-\alpha\omega^2\cos\omega t+\beta\omega\sin\omega t,\nonumber\\
\partial_t^3{z}(t)&=&\alpha\omega^3\sin\omega t+\beta\omega^2\cos\omega t.
\label{F10}
\end{eqnarray}
In this solution $I_{\rm STAT}$ obeys
\begin{equation}
\frac{2A(T)}{\gamma}I_{\rm STAT}=\omega\dot{z}^2(T)(\sin\omega T-\omega T\cos
\omega T)-2\omega^2z(T)\dot{z}(T)(1-\cos\omega T)+\omega^3 z^2(T)\sin \omega T.
\label{F11}
\end{equation}
Finally, we verify that this function is a solution to the Schr\"odinger
equation associated with (\ref{e34}) and identify the pre-factor as $A^{-1/2}(T
)$. The propagator is thus $A^{-1/2}(T)e^{iI_{\rm STAT}}$. Its covariant
generalization, obtained by using (\ref{e36}), is given in (\ref{e50}).

{}

\begin{thebibliography}{}
\bibitem{A9} Technically, what was shown in \cite{A8} is that the secular
equation that determines energy eigenvalues is real if and only if the
Hamiltonian is $\cP\cT$ symmetric. Since the solutions to an equation that is
not real cannot all be real, $\cP\cT$ invariance of a Hamiltonian is thus a
necessary condition for the reality of eigenvalues. The solutions to a real
equation can either be real or appear in complex conjugate pairs. Thus, $\cP\cT$
invariance alone does not guarantee reality. However, as noted in
Ref.~\cite{A8}, whether one has the real case or the complex-conjugate pair case
depends on whether or not the available $\cC$ operators of $\cP\cT$ theory
described in Ref.~\cite{A6} commute with $\cP\cT$. Commutation of both the
Hamiltonian and all available $\cC$ operators with $\cP\cT$ is then both
necessary and sufficient for the reality of energy eigenvalues.
\bibitem{A10a} While we do not discuss it here, we note that analogous to the
Hamiltonian given in (\ref{e15}), comparable results can be obtained for the
Hamiltonian $H=(1/2m)(p_\mu-ic_\mu)(p^\mu-ic^\mu)+m\omega^2x_\mu x^\mu/2$, where
$c^\mu$ is a real $\cP\cT$-odd four-vector.
\bibitem{A16} Transforming $H_{\rm PU}$ in (\ref{e34}) with $\cQ=i\alpha p_xp_z-
i\beta xz$ to $\tilde{H}=p_x^2/2\gamma+\gamma\omega_1^2x^2/2-p_z^2/2\gamma
\omega_1^2-\gamma\omega_1^2\omega_2^2z^2/2$, we see that if $p_z$ and $z$ were
Hermitian, the Hamiltonian would be unbounded below.
\bibitem{A16a} One can either represent the fourth-order derivative theory as a
quadratic theory based on two sets of conjugate position and momentum pairs or
as a quartic theory based on one conjugate pair alone. Both representations
yield the same energy eigenvalues.
\bibitem{A17} A relative minus sign can appear in (\ref{e31}) in the Hermitian
Hamiltonian case where there are no states with negative norm because as noted
in \cite{A13}, in this case the contour for the $p^0$ integration is not the
standard Feynman one. Specifically, in the generic form $1/(A+i\epsilon_a)-1/(B
+i\epsilon_b)=(B-A)/(AB+i\epsilon)$, where $\epsilon=B\epsilon_a+A\epsilon_b$
the sense of the contours depends on the values of $A$ and $B$.
\end{thebibliography}
\end{document}